\documentclass[pss]{wiley2sp} % provides pss two-column style
\usepackage{nicefrac}
\usepackage{amssymb}
\newcommand{\AVE}[1]{\left\langle{#1}\right\rangle}

\begin{document}

% Title of the article
\title{Topological phase transitions in bulk.}

% Abbreviated title for the page headers
\titlerunning{Topological phase transitions in bulk.}

% Authors
\author{%
  Stanislav Chadov\textsuperscript{\Ast,\textsf{\bfseries 1}},
  Janos Kiss\textsuperscript{\textsf{\bfseries 1}},
  J\"urgen K\"ubler\textsuperscript{\textsf{\bfseries 2}}, 
  Claudia Felser\textsuperscript{\textsf{\bfseries 1}}}

% Abbreviated list of authors for the page headers
\authorrunning{S.~Chadov et al.}

%E-mail-address of corresponding author
\mail{e-mail
  \textsf{stanislav.chadov@cpfs.mpg.de}, Phone:
  +49-0351-46462238, Fax: +49-0351-46463002}

% author's affiliations/addresses
\institute{%
  \textsuperscript{1}\,Max-Planck-Institut f\"ur Chemische Physik fester
  Stoffe, N\"othnitzer Str. 40, 01187 Dresden, Germany\\
  \textsuperscript{2}\,Institut f\"ur Festk\"orperphysik, Technische Universit\"at Darmstadt, 64289 Darmstadt, Germany}

\received{XXXX, revised XXXX, accepted XXXX} % do not change, will be filled in by the publisher
\published{XXXX} % do not change, will be filled in by the publisher

% Please select about four verbal keywords for your manuscript.
\keywords{topological insulator, random disorder, electron localization,  CPA}

\abstract{%
% This is a macro for the typesetting of two-column text in an
% abstract. It will typeset the two arguments in \abstcol{}{} as the
% left and right column inside the abstract box. At the
% columnbreak there will be always a columnbreak (\par), so both
% columns start with a new paragraph. No automatic column height
% balancing is done.
%
% If used with a \titlefigure it will silently output both
% parameters as consecutive paragraphs.
%
% The macro is defined exclusively inside the argument of \abstract{};
% if used outside it will raise an error.
%
% Usage: \abstcol{<left column>}{<right column>}
%\abstcol{%

We consider the analogy between the topological phase
transition which occurs as a function of spatial coordinate on a surface
of a non-trivial insulator, and the one which occurs in the bulk due to
the change of internal parameters (such as crystal field and
spin-orbit coupling). In both cases the system exhibits a Dirac cone,
which is the universal manifestation of topological phase transition,
independently on the type of driving parameters. In particular, this leads
to a simple way of determining the topological class based solely 
on the bulk information even for the systems with translational symmetry
broken by atomic disorder or by strong electron correlations. Here we
demonstrate this on example of the zinc-blende related semiconductors by
means of the {\it ab-initio} fully-relativistic band structure
calculations involving the coherent potential approximation (CPA) technique.
 }

% The class file requires the standard graphicx Latex package. See the 'LaTeX
% standard graphics and color packages documentation' for more information at
% <http://tug.ctan.org/tex-archive/macros/latex/required/graphics/grfguide.pdf>.
%
% Accepted figure file formats depend on which LaTeX flavour is used.
% Classic LaTeX is always able to use Encapsulated Postscript (EPS);
% PDFLaTeX can't use this but accepts PDF, JPG, PNG, and GIF formats.
%
% See examples for implementing graphics in floating figure environments later in this file.
% If \titlefigure is given, it takes as its mandatory parameter the
% name (without extension) of some figure file.

%\titlefigure[height=3.1cm]{empty2w}
%\titlefigurecaption{%
%  This is the caption of the \emph{optional} abstract figure. If
%  there is no abstract figure here, the abstract text should be divided
%  into both columns.
%}

\maketitle   % please do not remove

\section{Introduction.}

The remarkable ability of the semiconductors with topologically non-trivial band
structure (for the overveiw see  e.\,g. Ref.~\cite{HK10}) to exhibit the non-dissipative quantized surface spin-current attracts
great attention both in theoretical physics, material science and
spintronics engineering. This spin-current is represented by a pair of
opposite charge currents carrying opposite spin states. Each
charge current is induced by the intrinsic spin-orbit interaction, which
can be viewed as spin-dependent effective magnetic field. 
These effective magnetic fields quantize the electron orbits into the
Landau levels, which leads to a quantization of
the resulting spin-current in units of $(\nicefrac{1}{2}+n)\left(e^2/h\right)$. The absence of dissipation 
holds as long as the structure of the Landau levels is preserved which is
guaranteed by the time-reversal symmetry. In contrast to the bulk, the non-compensated spin-current can be observed
and detected on the surface, due to the space-reversal symmetry break
while keeping the time-reversal symmetry. On the surface the system exhibits the Dirac cone, which
can be observed directly by the  angular-resolved photoemission technique. 
This property of topological insulators is typically exploited in
the band structure calculations in order to probe the non-trivial nature
of the material. On the other hand, the reliable surface calculations
most often appear to be extremely resource demanding. Alternatively,
since the topological edge states emerge from the non-trivial
bulk band structure, it must be a way for determining the topological
class based on pure bulk information. Indeed, it is possible to
calculate the topological invariant Z$_2$ based on the parity analysis of the
occupied eigenstates~\cite{FKM07,FK07}. Obviously, this procedure cannot be
applied in situations when the eigenstates do not exist due to the
absence of the translational symmetry. The reasons for such symmetry break are
typically the non-periodic boundary conditions, atomic (chemical or structural) disorder, 
or the intrinsic electronic disorder caused by strong dynamical correlations. 
The hint is that in the first case, i.\,e. on the surface of
topological insulator one always observes the Dirac cone. At this moment it is
important to understand that the surface is not just a complicated
geometrical object with reduced dimensionality, potential step etc., but
it is a transition between different bulk phases. The parameter of this
transition is simply the spatial coordinate. In this context the  Dirac cone is not an ingenuous 
surface phenomenon, but is rather a fundamental manifestation of the 
transition between two distinct topological phases in certain parameters space. 
Indeed, the Dirac cone can occur in the bulk band structure as well: 
all materials which appear at the borderline between topologically non-trivial and trivial phases exhibit
the bulk Dirac cone~\cite{CQK+10}. This fact indicates that for
materials with the translational symmetry  broken by disorder 
the topologically non-trivial state still can  be determined based on the bulk information.

In the present work we would like to illustrate this fact by 
first-principles calculations taking into account the random disorder by
means of the Coherent Potential Approximation (CPA) alloy
theory~\cite{Sov67,But85}. Simultaneously, the CPA  serves as adiabatic transition
technique, which allows us to model the smooth transformation of one
material into an other and observe the details of the topological phase
transition in the bulk. This transformation has also a physical
relevance, i.\,e. it directly describes the effect of the random chemical substitution.
As a  real-case example we will consider the class of zinc-blende 
related structures, which contain a great manifold of non-trivial and
trivial semiconductors in which the details of their band structures are well-known.

\section{Topologically non-trivial systems with zinc-blende structure.}

\begin{figure}
\centering
\includegraphics[width=0.4\linewidth]{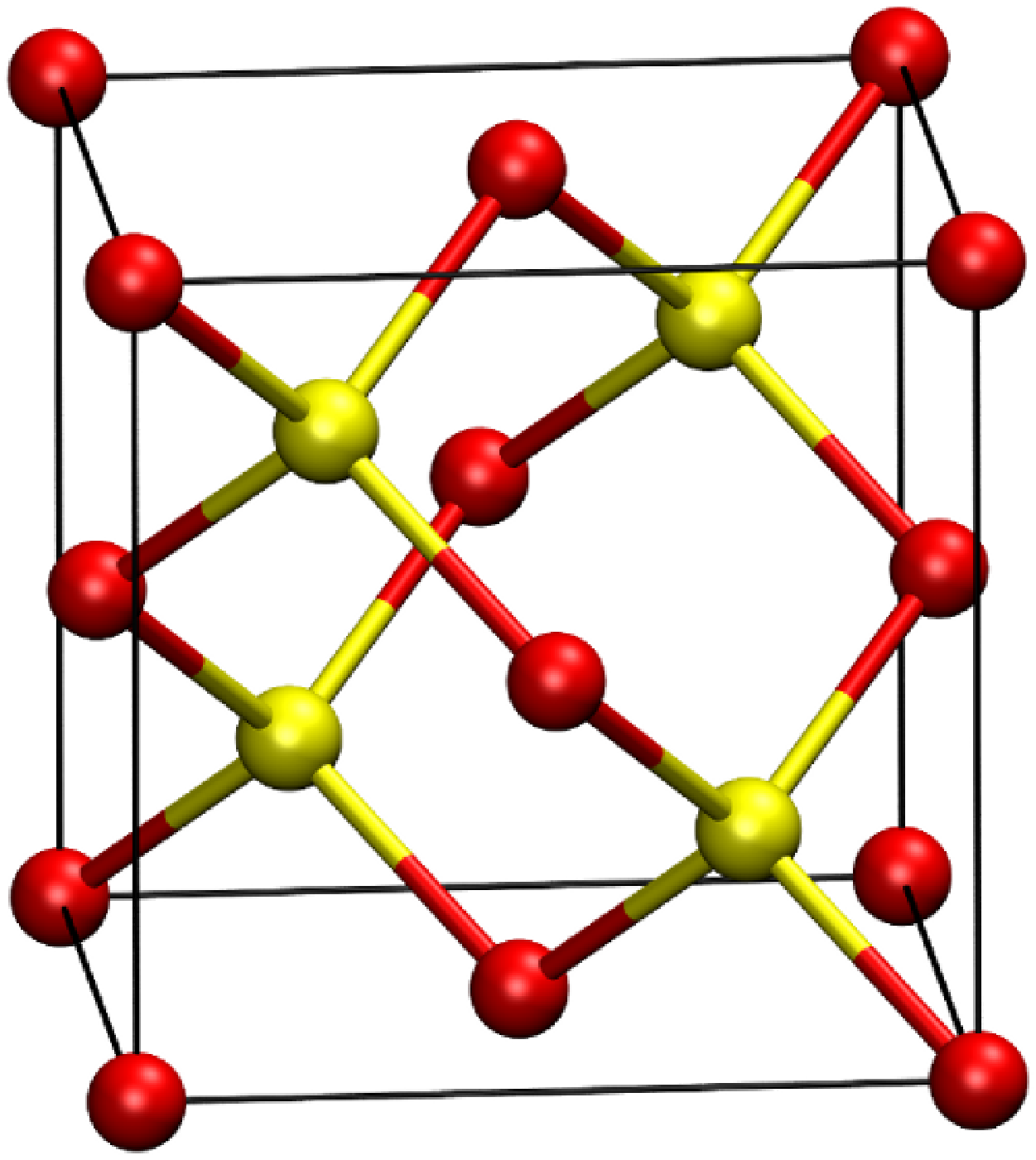}\hspace*{0.75cm}\includegraphics[width=0.4\linewidth]{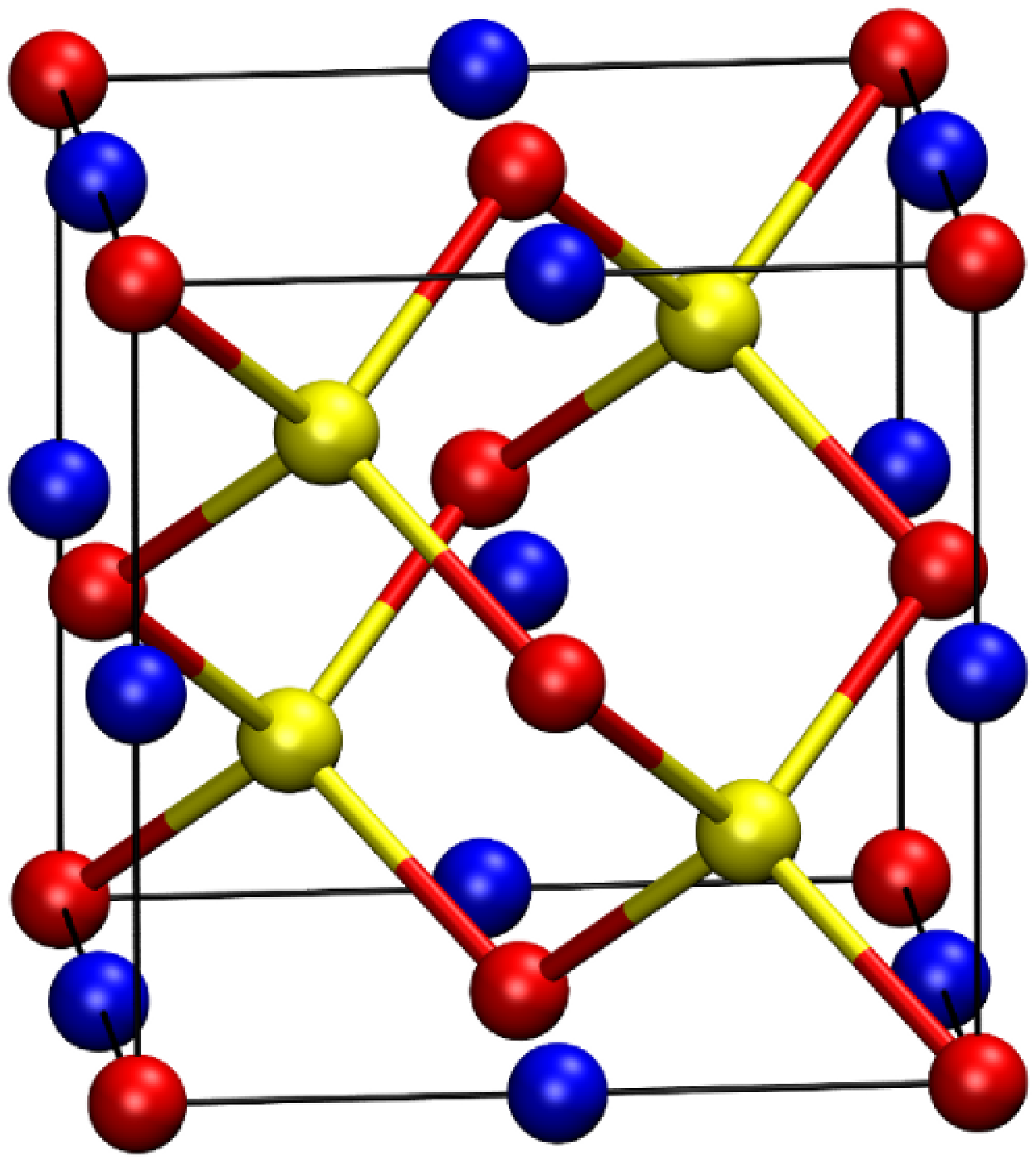}
\caption{(color online) Left: crystal structure of the zinc-blende binary YZ
  compound (space group No.~216). Yellow an red spheres represent the transition-metal Y and
  main-group elements Z. Right: by adding one more transition element X (blue
  sphere) one obtains ternary XYZ Heusler compound.\label{fig:zinc-blende-structure}}
\end{figure}

\begin{figure}
\centering
\includegraphics[width=1.0\linewidth]{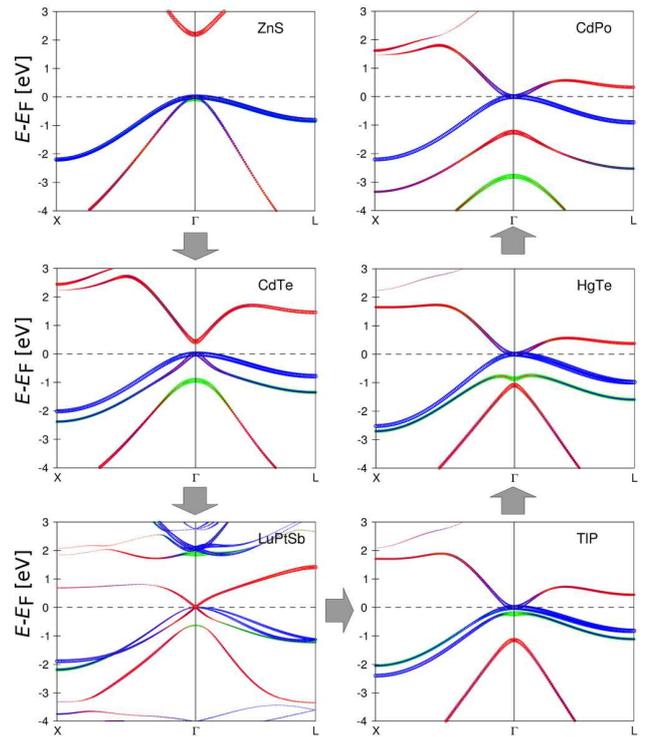}
\caption{(color online) Relativistic band structures calculated for the 
  ordered YZ binary zinc-blende (ZnS, CdTe, TlP, HgTe and CdPo) and related ternary XYZ Heusler (LuPtSb) materials. Red, green and blue
  colors refer to the orbital symmetries $s_{\nicefrac{1}{2}}$ (carried by the transition
  element Y), $p_{\nicefrac{1}{2}}$ and $p_{\nicefrac{3}{2}}$ (carried by the main group element Z), respectively. The gray arrows indicate
  the transition from trivial semiconductors ZnS, CdTe towards
  topological phase transition marked by the Dirac cone (LuPtSb), and
  further into the class of the non-trivial gapless semiconductors TlP,
  HgTe and CdPo. Calculations are performed by the PY-LMTO
  method~\cite{PYA}. As the   LDA~\cite{VWN80} form of the
  exchange-correlation potential was used, the band gaps of trivial
  semiconductors (ZnS, CdTe) are underestimated.
\label{fig:orbital-scheme}}
\end{figure}

Topological semimetals (or gapless semiconductors with topologically
non-trivial band structure) represent a rich and at the same time rather
simple class of materials, which allows to demonstrate the general
trends specific for the class of the 2D topological insulators as
a whole. The simplest systems in this class are the binaries with
zinc-blende structure (see Fig.~\ref{fig:zinc-blende-structure}) as
e.\,g. HgTe~\cite{BHZ06,KWB+07}. It is relatively easy to search for
the nonmagnetic semiconductors in this class by picking the pairs of
elements with average number of valence electrons per atom ${n_{\rm  val}=6}$~\cite{CQK+10,LWX+10}. 
%%%%%%%%%%%%%%%%%%%%%%%%%%%%%%%%
%%%%%%%%%%%%%%%%%%%%%%%%%%%%%%%%
Without restriction one can add here the whole
class of ternary Heusler semiconducting materials (so-called
half-Heusler semiconductors) which possess the same space group No.~216
and are different from binaries by having one more transition metal
element~\cite{CQK+10,LWX+10,SLM+10}. As it follows from these studies, 
all topologically-related physics of zinc-blende like compounds is
contained in the vicinity of the $\Gamma$-point: all trivial semiconductors of this type
exhibit a direct non-zero band gap, whereas all non-trivial materials are
gapless. It is surprising, but there was no exception found yet.
Here we also illustrate the difference between topological and trivial
materials within this class in terms of the band structure, by
giving few examples shown in Fig.~\ref{fig:orbital-scheme}.
%%%%%%%%%%%%%%%%%%%%%%%%%%%%%%%%
%%%%%%%%%%%%%%%%%%%%%%%%%%%%%%%%
The trivial band structure is characterized by the empty
$s_{\nicefrac{1}{2}}$ orbital of the transition metal atom Y and the fully filled
$p$-shell of the main-group element Z, split into
$p_{\nicefrac{3}{2}}$ and $p_{\nicefrac{1}{2}}$ by spin-orbit coupling. These compounds
exhibit a non-zero direct band-gap at the Fermi energy and are
represented by the light-element semiconductors with average nuclear number
$\AVE{Z}$ roughly less than 60: e.\,g. ZnS ($\AVE{Z}=23$), ZnSe~($\AVE{Z}=32$),
CdTe, ScPtSb ($\AVE{Z}=50$); with certain exceptions, as
e.\,g. LuNiBi (${\AVE{Z}\approx60.7}$)~\cite{CQK+10}.
In the non-trivial systems the $p_{\nicefrac{3}{2}}$ orbital is
half-filled, for this reason it sticks to the Fermi energy and within
cubic symmetry the system  becomes zero-gap semiconductor. At the same
time the $p_{\nicefrac{1}{2}}$ and $s_{\nicefrac{1}{2}}$ orbitals are fully occupied. This
so-called band inversion changes the parity of the wave function which is
directly related to the topological Z$_2$ invariant. Such materials are
typically heavy with ${\AVE{Z}\gtrsim 60}$: ZnPo (${\AVE{Z}=57}$), HgTe
and CdPo (${\AVE{Z}=66}$), YPtBi (${\AVE{Z}\approx66.7}$), LuPtBi (${\AVE{Z}\approx77.3}$)~\cite{CQK+10,LWX+10}; with certain exceptions,
as e.\,g. was found in present calculations for TlP (${\AVE{Z}=48}$).
Interesting case is represented by the borderline compounds, such as YPdBi, YPtSb
(${\AVE{Z}=56}$), or LuPtSb (${\AVE{Z}\approx66.7}$)~\cite{SLM+10}, which exhibit the Dirac cone in the bulk, i.\,e. can
be easily switched between topological or trivial states by small
variations of the stoichiometry or the lattice constant~\cite{CQK+10}. 

Obviously, there is also a continuum of alloys, which can be obtained by combining various
ordered zinc-blende (Y$_{1-y}$Y$'_y$)(Z$_{1-z}$Z$'_z$) or half-Heusler
materials (X$_{1-x}$X$'_x$)(Y$_{1-y}$Y$'_y$)(Z$_{1-z}$Z$'_z$). 
By varying the ratios of  elements allows to tune the properties of the system continuously in a
 wide range. For example, by substituting Hg with Cd in
 (Hg$_{1-x}$Cd$_x$)Te series one obtains the zero-gap materials
 for ${x\lesssim0.17}$ and non-zero band gap for
 ${x\gtrsim0.83}$~\cite{CXT83}, which correspond to the non-trivial and
 trivial regimes, respectively.

\section{Chemical disorder in terms of the CPA alloy theory.}

In the context of the first-principles description of the electronic structure,
the most practical way to treat the random disorder is by using the so-called Coherent Potential Approximation
(CPA)~\cite{Sov67,But85}. In fact, until now the CPA (and its extensions) remains the only
 technique which incorporates the effects of the energy-dependent shift and lifetime broadening -- the essential 
features of the electron localization caused by chemical disorder, which
are not accessible within other theories, as e.\,g. the VCA~\cite{Nor31} or  supercell calculations. 

The idea of CPA can be sketched by considering the  example of 
binary  alloy A$_{x}$B$_{1-x}$ (${0\le x\le1}$), with each lattice site occupied either by
 atomic sorts A or B with probabilities $x$ or $1-x$, respectively. For
 simplicity we let the atomic potential to be characterized by a single
 energy level, $\epsilon_{\rm A}$ and $\epsilon_{\rm B}$, for each sort respectively.
In the second quantization form the Hamiltonian of A$_{x}$B$_{1-x}$ alloy $H_{\rm AB}$ can be written as 
\begin{equation}
H_{\rm AB}=\sum_{k}\epsilon_k a_k^{\dagger}a_k^{} +
\sum_{i}\epsilon_ia_i^{\dagger}a_i\,.
\end{equation}
The first term represents the sum over the valence states indexed by their
wave vectors $k$ with kinetic energies $\epsilon_{k}$. The second term
is the sum over the lattice sites $i$; $\epsilon_{i}$ is the randomly
fluctuating variable, equal to $\epsilon_{\rm A}$ or  $\epsilon_{\rm B}$
with probabilities $x$ or $1-x$, respectively. 

\begin{figure}
\centering
\begin{minipage}{0.35\linewidth}
  \centering 
  A$_x$B$_{1-x}$
\end{minipage}~~~~
\begin{minipage}{0.35\linewidth}
  \centering 
  C
\end{minipage}

\vspace*{1ex}

\includegraphics[width=0.75\linewidth]{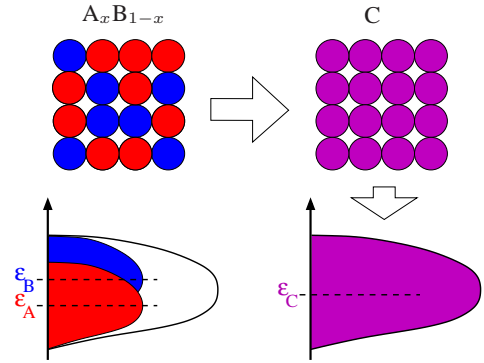}
\caption{(color online) Mean-field treatment of random disorder. 
  Random binary A$_{x}$B$_{1-x}$ alloy (A and B atoms are shown as red and
  blue spheres)  is substituted by the regular array of the single effective
  atomic sort C (purple spheres). The effective potential
  $\epsilon_{\rm C}$ is adjusted in order to exhibit the same properties
  as the original disordered binary: for example, the same density of states.\label{fig:meanfield}}
\end{figure}

The straightforward solution for such Hamiltonian represents a complicated many-body
problem. To make it practically accessible, one needs to find an
adequate mean-field approximation, i.\,e. to substitute the real
disordered system A$_{x}$B$_{1-x}$ by the effective regular system C 
\begin{eqnarray}
H_{\rm AB}\approx H_{\rm C} = \sum_{k}\epsilon_k a_k^{\dagger}a_k^{} +
\sum_{i}\epsilon_{\rm C} a_i^{\dagger}a_i \nonumber \\ = \sum_{k}\epsilon_k a_k^{\dagger}a_k^{} +
\epsilon_{\rm C}\sum_{i}a_i^{\dagger}a_i\,,
\end{eqnarray}
which emulates the same properties. This idea is sketched on
Fig.~\ref{fig:meanfield}.
%%%%%%%%%%%%%%%%%%%%%%%%%%%%%%%%
%%%%%%%%%%%%%%%%%%%%%%%%%%%%%%%%
The quality of such emulation is a matter of how the effective potential
$\epsilon_{\rm C}$ is adjusted. It is easy to show, for example, that
the straightforward averaging over the single-site potentials 
misses much of important physics. Indeed, since both $\epsilon_{\rm A}$
and $\epsilon_{\rm B}$ are the real-valued quantities, the effective
regular potential ${\epsilon_{\rm C}=x\cdot\epsilon_{\rm A}+(1-x)\cdot\epsilon_{\rm B}}$
will be real-valued as well. Thus, the solutions of the corresponding
Schr\"odinger equation will be the delocalized Bloch waves which propagate without
scattering and always lead to an infinite residual conductivity in case
of a metal. 

The CPA is based on a  more rational choice of the mean-field variable by using the
one-particle Green's function $G_{\rm AB}$. Since it represents a response of the
whole system to the $\delta$-like perturbation,
i.\,e. 
\begin{eqnarray}
\left[\epsilon-H_{\rm AB}(\vec r)\right]G_{\rm AB}(\vec r,\vec r',\epsilon) = \delta(\vec r-\vec r')\,, 
\end{eqnarray}
it can be seen as a weighted superposition of responses from all 
configurations $\Omega$ present in the system:
\begin{eqnarray}
G_{\rm AB} = \sum_{\Omega} x_{\Omega}G_{\Omega}\,,\ \ \ \sum_{\Omega}x_{\Omega}=1\,.
\end{eqnarray}
Each configuration $\Omega$ can be presented as a perturbation of the effective
regular system C, i.\,e. it can be defined using the Dyson equation: 
\begin{eqnarray}
G_{\Omega} &=& G_{\rm C}+G_{\rm
  C}\sum_i\left(\epsilon_{i,\Omega}-\epsilon_{\rm   C}\right)G_{\Omega}\nonumber \\
&=& G_{\rm C}+G_{\rm C}T_{\Omega}G_{\rm C}\,,
\label{eqv:configurations}
\end{eqnarray}
where $\epsilon_{i,\Omega}$ is equal $\epsilon_{\rm A}$ or
$\epsilon_{\rm B}$, depending on the configuration $\Omega$; the
perturbation with respect to the regular system C is fully described by the term
\begin{eqnarray}
T_{\Omega} &=& \sum_{i}\left(\epsilon_{i,\Omega}-\epsilon_{\rm C}\right) +
\nonumber \\ &+&\sum_{i,j}\left(\epsilon_{i,\Omega}-\epsilon_{\rm
  C}\right)G_{\rm C}\left(\epsilon_{j,\Omega}-\epsilon_{\rm C}\right) +
\ldots \nonumber \\ && \hspace*{0.5cm} =\displaystyle
\frac{\sum_{i}\left(\epsilon_{i,\Omega}-\epsilon_{\rm
    C}\right)}{1-\sum_{i}\left(\epsilon_{i,\Omega}-\epsilon_{\rm
    C}\right)G_{\rm C}}\,.
\label{eqv:tau}
 \end{eqnarray}
Thus, the requirement 
\begin{eqnarray}
G_{\rm AB} = G_{\rm C}
\end{eqnarray}
is equivalent to the cancellation of all partial perturbations in a total sum
\begin{eqnarray}
  \sum_{\Omega} x_{\Omega}T_{\Omega} = 0\,,
\label{eqv:cancellation}
\end{eqnarray}
which completes the self-consistent set of equations, needed to
determine $\epsilon_{\rm C}$. Since the definition~(\ref{eqv:tau})
involves the Green's function $G_{\rm C}$, which is complex-valued and
energy-dependent, the effective potential $\epsilon_{\rm C}$ appears to
be complex-valued and energy-dependent as well. Its real part
Re\,$\epsilon_{\rm C}$ defines the energy position of the
electronic state, whereas the imaginary part Im\,$\epsilon_{\rm C}$ is responsible for
its finite lifetime broadening caused by disorder. The latter is
the necessary ingredient for the calculation of residual conductivity
(except of the Berry phase contribution~\cite{XCN10} which is a pure band structure effect).
At the same time, the effective system  formally remains to be translational
invariant, thus once $\epsilon_{\rm C}$ is determined, we can restrict
all further calculations to the primitive unit cell. 

\begin{figure}
\centering
\includegraphics[width=0.9\linewidth]{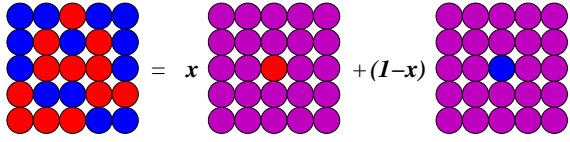}
\caption{(color online) Single-site CPA. Random binary A$_{1-x}$B$_x$ alloy (A and B atoms are shown as red and
  blue spheres)  is approximated as a superposition of two random impurity
  Green's functions (the impurities of types A and B are embedded in the
  regular host of the effective atoms C shown in purple) weighted with corresponding concentrations $x$ and ${1-x}$. \label{fig:cpa}}
\end{figure}

In order to simplify the set of the mean-field equations
(\ref{eqv:configurations}, \ref{eqv:tau}, \ref{eqv:cancellation}) we
assume further that the only important configurations are those where
the atoms are placed fully randomly, i.\,e. without certain nearest
environment preference. In this case, only two configurations are left:
random single impurities of type A  and of type B, embedded in the C host (see Figure~\ref{fig:cpa}). 
%%%%%%%%%%%%%%%%%%%%%%%%%%%%%%%%
%%%%%%%%%%%%%%%%%%%%%%%%%%%%%%%%
Thus, the shortcoming of the single-site CPA is the absence of local environment effects. The latter
can be, however, taken into account by its non-local extensions~\cite{MSC83,VB00,RSG03}.
On the other hand, for the case of random isovalent substitution, the single-site CPA remains a quite reasonable approximation even in the
diluted limit, since the isovalent atoms intermix without extra environmental
preference. The only condition is to preserve the zinc-blende structure which
is again guaranteed by isovalency.

In the following all electronic structure calculations
were  carried out using the fully-relativistic
Korringa-Kohn-Rostoker (KKR) Green's function method implemented within the 
SPR-KKR package~\cite{EKM11} employing the density functional theory
framework. Due to the multiple-scattering construction of the Green's function~\cite{FS80}, 
the method provides a suitable base for mean-field approaches like CPA. 
Since the electronic structure of disordered materials in general does not
show well defined Bloch eigenstates, their electronic structure is described
by the Bloch-spectral function (BSF) defined by the imaginary part
of the alloy Green's function that is diagonal in momentum space~\cite{FS80}.
The exchange and correlation was treated using the Vosko-Wilk-Nusair
form of the local density approximation (LDA)~\cite{VWN80}. 
LDA is known for its typical underestimation of the band gaps of semiconductors.
However, in case of our study this issue is not critical, since the observed 
trends are qualitatively well described even with LDA.

\section{Topological bulk phase transition induced by chemical substitution.}

\begin{figure}
\centering
\includegraphics[width=0.8\linewidth]{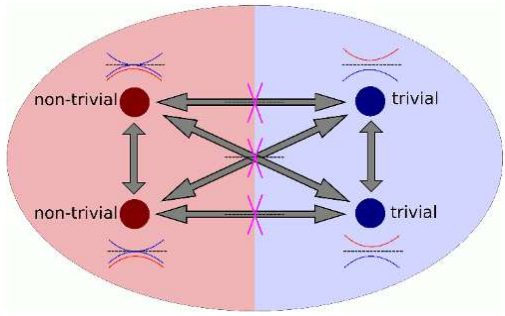}
\caption{(color online) Schematic illustration of the topological phase
  transition via substitution. Light red and blue areas distinguish
  between manifolds of topologically non-trivial and trivial materials,
  respectively. The corresponding band structures are schematically
  given in the vicinity of the red and blue spheres which indicate
  certain compounds from topological and trivial manifolds, respectively. The arrows indicate the transformation of 
   one material into the other, e.\,g. by chemical substitution or by
    overgoing through the interface. The Dirac cone indicated by light
   purple lines always appears for the borderline case. \label{fig:adiabatic_path}}
\end{figure}

In the following we will track the evolution of the BSF driven by random
chemical substitution along three distinct adiabatic paths, as
depicted in Fig.~\ref{fig:adiabatic_path}, first by going between the
pair of trivial, second - between the pair of non-trivial  materials. In
the third case we span the path from the non-trivial to the trivial one
in order to study the details of the topological phase transition.
%%%%%%%%%%%%%%%%%%%%%%%%%%%%%%%%
%%%%%%%%%%%%%%%%%%%%%%%%%%%%%%%%
In order to keep the conditions of adiabatic transition we perform
the random substitution in isovalent manner, by keeping the chemical
bonding rules and the crystal structure: i.\,e. YZ~$\rightarrow$~(Y$_{1-x}$Y$'_x$)(Z$_{1-x}$Z$'_x$)~$\rightarrow$~Y$'$Z$'$.
For the alloy series the  lattice constant is approximated linearly
between the experimental constants for the pure materials.

\begin{figure*}
\centering
\includegraphics[width=0.95\linewidth]{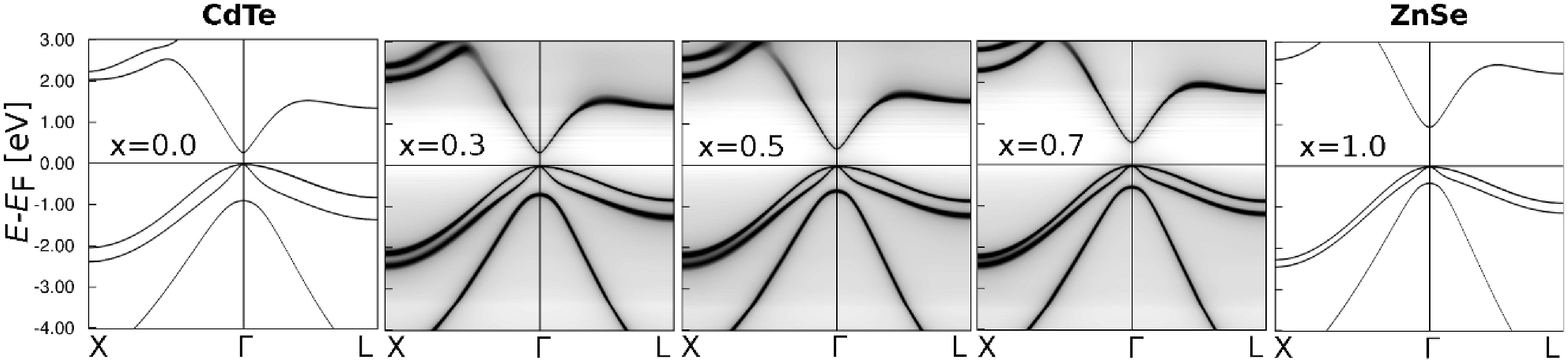}
\caption{ Bloch-spectral functions calculated for the series of alloys
  (Cd$_{1-x}$Zn$_x$)(Te$_{1-x}$Se$_x$), ${x = 0, 0.3, 0.5, 0.7, 1}$. 
  Since the pure compounds CdTe (${x=0}$) and ZnSe (${x=1}$) belong to
  the same  topological (trivial) class, all intermediate compositions exhibit similar
  Bloch-spectral functions. \label{fig:CdTe-ZnSe}}
\end{figure*}

As an example for the first case, we select two trivial materials,
namely CdTe and ZnSe which are well-known semiconductors with non-zero
band gaps. Both systems exhibit the same band structure features (Fig.~\ref{fig:CdTe-ZnSe}):
the conduction band is formed by the $s_{\nicefrac{1}{2}}$-symmetry
($5s$ for Cd and $4s$ for Zn), whereas the top of the valence band carries
the $p_{\nicefrac{3}{2}}$ symmetry ($5p$ and $4p$ orbitals for Te and Se, respectively).
%%%%%%%%%%%%%%%%%%%%%%%%%%%%%%%%
%%%%%%%%%%%%%%%%%%%%%%%%%%%%%%%%
As it follows from the BSFs for the  intermediate compositions,
the random substitution of Cd by isovalent Zn and Te by isovalent Se
does not change the energy distribution of the orbital characters, thus
all the alloys in between exhibit  similar band
structures. On the other hand, one can easily notice their principle
difference from the ordered materials: for ${0<x<1}$ the symmetry
analysis of eigenvalues cannot be applied directly, since
there are no pure Bloch eigenstates, i.\,e. in the disordered
regime the $\delta$-like poles of the BSF turn into overlapping
Lorentzians, spread over the whole $k$-space. Despite that we still can trace the evolution of their maxima in analogy to dispersion relation for the pure
material, the only strict fact which tells us that all these materials 
belong to the same topological class is the absence of the Dirac-like
dispersion within these series.

\begin{figure*}
\centering
\includegraphics[width=0.95\linewidth]{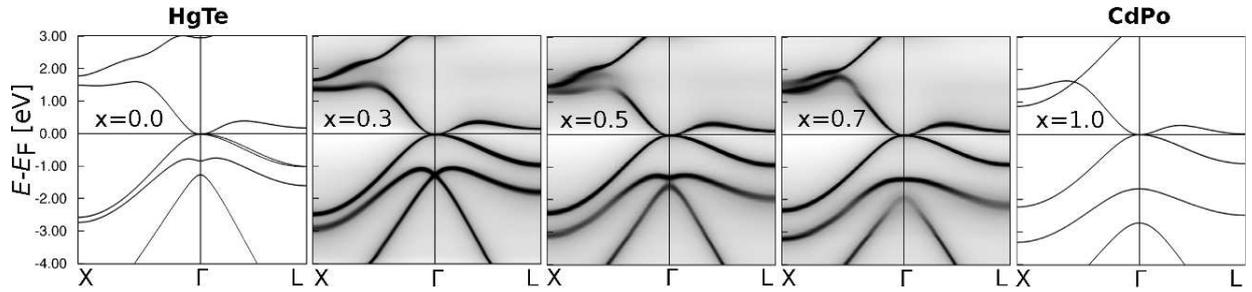}
\caption{ Bloch-spectral functions calculated for the series of alloys
  (Hg$_{1-x}$Cd$_x$)(Te$_{1-x}$Po$_x$), ${x = 0, 0.3, 0.5, 0.7, 1}$. 
  Since the pure compounds HgTe (${x=0}$) and CdPo (${x=1}$) belong to
  the same  topological (non-trivial) class, all intermediate
  compositions exhibit similar gapless  Bloch-spectral functions. \label{fig:HgTe-CdPo}}
\end{figure*}

In the second case we transform the topologically non-trivial HgTe into
another non-trivial zinc-blende system CdPo. The Po zinc-blende compounds
 (CdPo, ZnPo) were not considered earlier in the context of topology, however their
 crystal and band structures are well-known long
 ago~\cite{Mad82,BZF07}. As the presently calculated band structures show, CdPo is a gapless
 semiconductor with the bands alignment equivalent to HgTe (see
 Fig.~\ref{fig:orbital-scheme}).  As it follows from  the calculated
 series shown on Fig.~\ref{fig:HgTe-CdPo}, 
%%%%%%%%%%%%%%%%%%%%%%%%%%%%%%%%
%%%%%%%%%%%%%%%%%%%%%%%%%%%%%%%%
the situation is in principle very similar to the first  case, with only difference that all materials belong to the non-trivial
 topological class, i.\,e. (Hg$_x$Cd$_{1-x}$)(Te$_{x}$Po$_{1-x}$), (${0<x<1}$) remain gapless
semiconductors and none of them exhibits the Dirac cone. On the other
hand, it is interesting to admit, that due to different energy placements of the
$s$-state in pure HgTe and CdPo  compounds one observes the cone-like
feature for the composition ${x\approx0.3}$ at about -1.5~eV. Indeed, in case of HgTe the $s$-state
is situated below both $p_{\nicefrac{1}{2}}$ and $p_{\nicefrac{3}{2}}$
and in case of CdPo  - right in between (see Fig.~\ref{fig:orbital-scheme}). By substituting Te to Po, the effective
spin-orbit split increases by pushing the $p_{\nicefrac{1}{2}}$-character down in
energy until it interchanges its position with $s$-character state at
${x\approx0.3}$. Since this re-ordering of the states is not connected
with their re-occupation, the  parity of the total wave function does
not change and the rest of the alloy series remains in the same topological class.

In the last case the adiabatic path is going from the non-trivial
 CdPo to the trivial ZnSe semiconductor. As it follows from BSFs
 calculated
%%%%%%%%%%%%%%%%%%%%%%%%%%%%%%%%
\begin{figure*}
\centering
\includegraphics[width=0.95\linewidth]{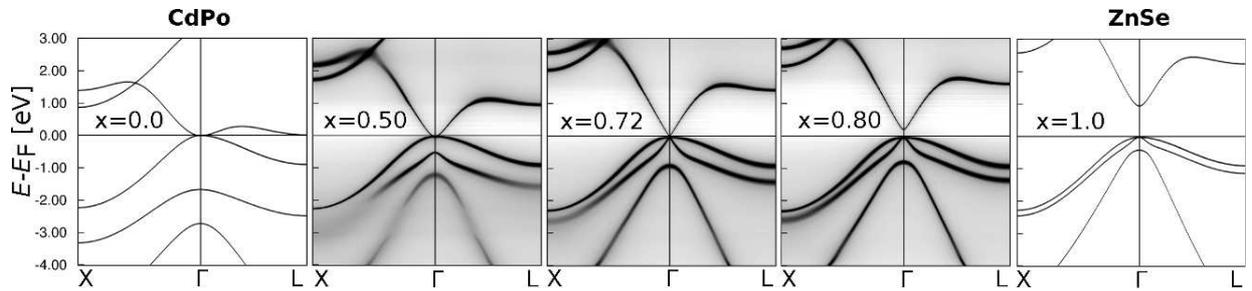}
\caption{ Bloch-spectral functions calculated for the series of alloys
  (Cd$_{1-x}$Zn$_x$)(Po$_{1-x}$Se$_x$), ${x = 0, 0.5, 0.72, 0.8,
    1}$. Since the pure compounds CdPo (${x=0}$) and ZnSe (${x=1}$) belong to
  different topological classes, there exists a certain intermediate 
  composition which exhibits the topological phase transition state
  (${x\approx 0.72}$). All compositions ${x<0.72}$ are topologically
  non-trivial and characterized by zero-gap, whereas ${x>0.72}$ are
  trivial and exhibit a real band gap, which increases towards
  ${x=1}$. The composition at ${x=0.72}$ is characterized by the clear Dirac
  cone centered at the Fermi energy. 
\label{fig:CdPo-ZnSe}}
\end{figure*}
%%%%%%%%%%%%%%%%%%%%%%%%%%%%%%%%
for the series (Cd$_{1-x}$Zn$_x$)(Po$_{1-x}$Se$_x$) (Fig.~\ref{fig:CdPo-ZnSe}),  we indeed encounter certain
composition (${x_{\rm c}\approx 0.72}$) characterized by the bulk Dirac
cone. All other alloys with ${x<0.72}$ are gapless semiconductors and 
 topologically non-trivial,  whereas  those for ${x>0.72}$ are in the trivial regime,
with the non-zero band gap. We briefly describe the most important
steps of the BSF evolution  in the vicinity of the $\Gamma$ point: the Po $6p$-band is split due to spin-orbit
coupling into a $p_{\nicefrac{3}{2}}$-state (of $\Gamma_8$ symmetry in
the pure case)  sitting right at the Fermi energy and a
$p_{\nicefrac{1}{2}}$-state (of $\Gamma_7$ symmetry) which is about 3~eV lower. The Cd
$5s_{\nicefrac{1}{2}}$ band (of $\Gamma_6$ symmetry) is placed by the crystal field in between, at about
-2~eV.  Since by substitution we effectively decrease the spin-orbit
coupling and increase the crystal field, both $p_{\nicefrac{1}{2}}$ and
$s_{\nicefrac{1}{2}}$-like states rise in energy by shifting towards the
Fermi level. At the ``critical'' composition ${x_{\rm c}=0.72}$ the
system undergoes a topological phase transition manifested by the
Dirac cone formed by a mixture of $p$- and $s$-symmetric states. 

\begin{figure*}
\centering
\includegraphics[width=0.95\linewidth]{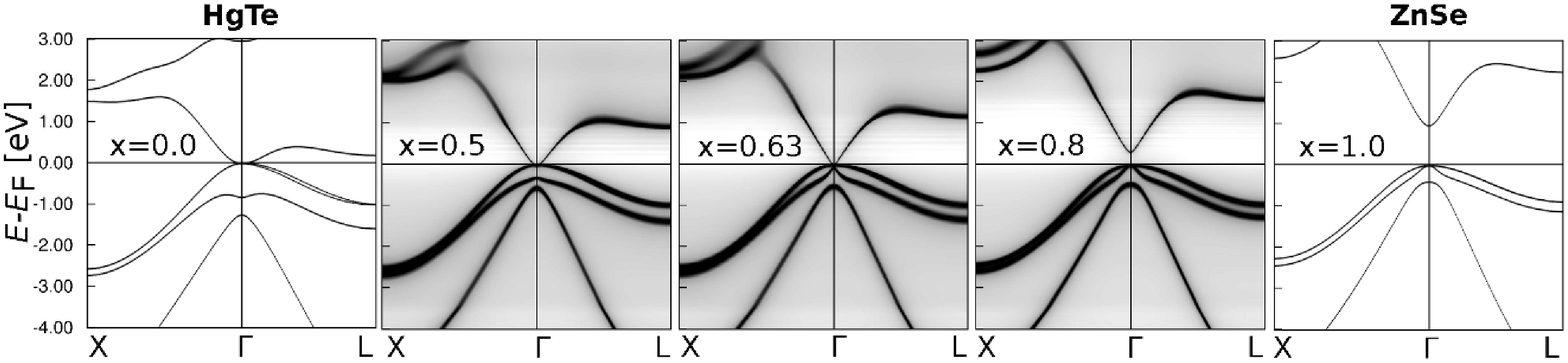}
\caption{ Bloch-spectral functions calculated for the series of alloys
  (Hg$_{1-x}$Zn$_x$)(Te$_{1-x}$Se$_x$), ${x = 0, 0.5, 0.72, 0.8,
    1}$. Since the pure compounds HgTe (${x=0}$) and ZnSe (${x=1}$) belong to
  different topological classes, there exists a certain intermediate 
  composition which exhibits the topological phase transition state
  (${x\approx 0.63}$). All compositions ${x<0.63}$ are topologically
  non-trivial and characterized by zero-gap, whereas ${x>0.63}$ are
  trivial and exhibit a real band gap, which increases towards
  ${x=1}$. The composition at ${x=0.63}$ is characterized by the clear Dirac
  cone centered at the Fermi energy. 
\label{fig:HgTe-ZnSe}}
\end{figure*}

Due to the extremely deep energy position of the Cd $s$-shell in CdPo
compound (approximately at -2~eV), the topological phase transition occurs within rather ZnSe-rich regime
${x_{\rm c}=0.72 > 0.5}$, which is not so severely disordered as in case
of $x=0.5$. Indeed, as it follows from the similar adiabatic path connecting
non-trivial HgTe with trivial ZnSe semiconductor (see Fig.~\ref{fig:HgTe-ZnSe}), the borderline
concentration is shifted towards non-trivial regime since the
$s$-symmetric state of the pure non-trivial alloy component is situated
closer to the Fermi energy.
%%%%%%%%%%%%%%%%%%%%%%%%%%%%%%%%
%%%%%%%%%%%%%%%%%%%%%%%%%%%%%%%%
Despite that the borderline composition is then shifted closer to the
maximally disordered system ${x_{\rm c}=0.63\approx 0.5}$, the Dirac
cone exhibits rather well-defined Bloch-like states compared to the rest
of the band structure.

Obviously, that the edge states with linear dispersion at the surface of
a topological insulator  and the states forming the Dirac cone  in the
bulk manifest the same transition phenomenon. The qualitative difference is only
that due to the remaining effective translational symmetry (the difference is
that the electrons are localized by disorder and are not Bloch-like), we observe two replica of Dirac cones with opposite
spins on top of one another, whereas due to the break of space-reversal 
symmetry at the surface only a single cone remains, which exposes the
adiabatic spin-current.

\acknowledgement

Financial support by the DFG project FOR 1464
“ASPIMATT” (1.2-A) is gratefully acknowledged.

\providecommand{\WileyBibTextsc}{}
\let\textsc\WileyBibTextsc
\providecommand{\othercit}{}
\providecommand{\jr}[1]{#1}
\providecommand{\etal}{~et~al.}

\bibliographystyle{pss}
%\bibliography{database}

\begin{thebibliography}{[10]}

\bibitem{HK10}% article
 \textsc{M.\,Z. Hasan} and  \textsc{C.\,L. Kane}\iffalse \textit{Colloquium} :
  Topological insulators\fi,
 \jr{Rev. Mod. Phys.} \textbf{82}, 3045--3067 (2010).


\bibitem{FKM07}% article
 \textsc{L.~Fu},  \textsc{C.\,L. Kane},  and  \textsc{E.\,J. Mele}\iffalse
  Topological insulators in three dimensions\fi,
 \jr{Phys. Rev. Lett.} \textbf{98}, 106803 (2007).


\bibitem{FK07}% article
 \textsc{L.~Fu} and  \textsc{C.\,L. Kane},
 \jr{Phys. Rev. B} \textbf{76}, 45302 (2007).


\bibitem{CQK+10}% article
 \textsc{S.~Chadov},  \textsc{X.\,L. Qi},  \textsc{J.~K\"ubler},
  \textsc{G.\,H. Fecher},  \textsc{C.~Felser},  and  \textsc{S.\,C.
  Zhang}\iffalse \textrm{Tunable multifunctional topological insulators in
  ternary Heusler compounds}\fi,
 \jr{Nature Materials} \textbf{9}, 541–--545 (2010).


\bibitem{Sov67}% article
 \textsc{P.~Soven}\iffalse \textrm{Coherent potential model of substitutional
  disordered alloys}\fi,
 \jr{Phys. Rev.} \textbf{156}, 809--813 (1967).


\bibitem{But85}% article
 \textsc{W.\,H. Butler}\iffalse \textrm{Theory of electronic transport in
  random alloys: Korringa-Kohn-Rostoker coherent-potential approximation}\fi,
 \jr{Phys. Rev. B} \textbf{31}, 3260--3277 (1985).


\bibitem{BHZ06}% article
 \textsc{B.\,A. Bernevig},  \textsc{T.\,L. Hughes},  and  \textsc{S.\,C.
  Zhang}\iffalse Quantum spin hall effect and topological phase transition in
  hgte quantum wells\fi,
 \jr{Science} \textbf{314}, 1757 (2006).


\bibitem{KWB+07}% article
 \textsc{M.~K\"onig},  \textsc{S.~Wiedmann},  \textsc{C.~Br\"une},
  \textsc{A.~Roth},  \textsc{H.~Buhmann},  \textsc{L.~Molenkamp},
  \textsc{X.\,L. Qi},  and  \textsc{S.\,C. Zhang}\iffalse Quantum spin hall
  insulator state in hgte quantum wells\fi,
 \jr{Science} \textbf{318}, 766 (2007).


\bibitem{LWX+10}% article
 \textsc{H.~Lin},  \textsc{L.\,A. Wray},  \textsc{Y.~Xia},  \textsc{S.~Xu},
  \textsc{S.~Jia},  \textsc{R.\,J. Cava},  \textsc{A.~Bansil},  and
  \textsc{M.\,Z. Hasan}\iffalse \textrm{Half-Heusler ternary compounds as new
  multifunctional experimental platforms for topological quantum phenomena}\fi,
 \jr{Nature Materials} \textbf{9}, 546–--549 (2010).


\bibitem{SLM+10}% article
 \textsc{W.~Al-Sawai},  \textsc{H.~Lin},  \textsc{R.\,S. Markiewicz},
  \textsc{L.\,A. Wray},  \textsc{Y.~Xia},  \textsc{S.\,Y. Xu},  \textsc{M.\,Z.
  Hasan},  and  \textsc{A.~Bansil},
 \jr{Phys. Rev. B} \textbf{82}, 125208 (2010).


\othercit
\bibitem{PYA}% inbook
 \textsc{A.~Perlov},  \textsc{A.~Yaresko},  and  \textsc{V.~Antonov},
\textrm{Spin-polarized Relativistic Linear Muffin-tin Orbitals Package for
  Electronic Structure Calculations, PY-LMTO.},
unpublished).


\bibitem{VWN80}% article
 \textsc{S.\,H. Vosko},  \textsc{L.~Wilk},  and  \textsc{M.~Nusair}\iffalse
  Accurate spin-dependent electron liquid correlation energies for local spin
  density calculations: a critical analysis\fi,
 \jr{Can. J. Phys.} \textbf{58}, 1200 (1980).


\bibitem{CXT83}% article
 \textsc{J.~Chu},  \textsc{S.~Xu},  and  \textsc{D.~Tang},
 \jr{Appl. Physics Lett.} \textbf{43}, 1064 (1983).


\bibitem{Nor31}% article
 \textsc{L.~N{\o}rdheim},
 \jr{Ann.~Phys.~(Leipzig)} \textbf{9}, 607 (1931).


\bibitem{XCN10}% article
 \textsc{D.~Xiao},  \textsc{M.\,C. Chang},  and  \textsc{Q.~Niu},
 \jr{Rev. Mod. Phys.} \textbf{82}, 1959--2007 (2010).


\bibitem{MSC83}% article
 \textsc{A.~Mookerjee},  \textsc{V.\,K. Srivastavat},  and
  \textsc{V.~Choudhry}\iffalse \textrm{Electronic structure of disordered
  alloys: II. Self-consistent cluster CPA. Application to III-V ternary
  alloys}\fi,
 \jr{J. Phys. C: Solid State Phys.} \textbf{16}, 4555--4564 (1983).


\bibitem{VB00}% article
 \textsc{C.\,I. Ventura} and  \textsc{R.\,A. Barrio}\iffalse
  \textrm{Self-consistent cluster {CPA} methods and the nested {CPA}
  theory}\fi,
 \jr{Physica B} \textbf{281-282}, 855--856 (2000).


\bibitem{RSG03}% article
 \textsc{D.\,A. Rowlands},  \textsc{J.\,B. Staunton},  and  \textsc{B.\,L.
  Gy\"orffy}\iffalse \textrm{Korringa-Kohn-Rostoker nonlocal coherent-potential
  approximation}\fi,
 \jr{Phys. Rev. B} \textbf{67}, 115109 (2003).


\bibitem{EKM11}% article
 \textsc{H.~Ebert},  \textsc{D.~K\"odderitzsch},  and
  \textsc{J.~Min\'{a}r}\iffalse \textrm{Calculating condensed matter properties
  using the KKR-Green's function method\ -\ recent developments and
  applications}\fi,
 \jr{Rep. Prog. Phys.} \textbf{74}(9), 096501 (2011).


\bibitem{FS80}% article
 \textsc{J.\,S. Faulkner} and  \textsc{G.\,M. Stocks}\iffalse Calculating
  properties with the coherent-potential approximation\fi,
 \jr{Phys. Rev. B} \textbf{21}, 3222 (1980).


\othercit
\bibitem{Mad82}% inbook
 \textsc{O.~Madelung} (ed.),
Semiconductors, Physics of II–VI and I–VII Compounds, Semi-magnetic
  Semiconductors, Landolt-B\"ornstein, New Series,  Vol.\,III/17a,
 (Springer, Berlin, 1982).


\bibitem{BZF07}% article
 \textsc{A.~Boukra},  \textsc{A.~Zaoui},  and  \textsc{M.~Ferhat},
 \jr{Solid State Commun.} \textbf{141}, 523--528 (2007).


\end{thebibliography}

\end{document}